# Manufacturing and Testing of Accelerator Superconducting Magnets

*L. Rossi[1]*
CERN, Geneva, Switzerland

**Abstract**
Manufacturing of superconducting magnet for accelerators is a quite complex process that is not yet fully industrialized. In this paper, after a short history of the evolution of the magnet design and construction, we review the main characteristics of the accelerator magnets having an impact on the construction technology. We put in evidence how the design and component quality impact on construction and why the final product calls for a total-quality approach. LHC experience is widely discussed and main lessons are spelled out. Then the new $Nb_3Sn$ technology, under development for the next generation magnet construction, is outlined. Finally, we briefly review the testing procedure of accelerator superconducting magnets, underlining the close connection with the design validation and with the manufacturing process.

## 1    Introduction

The manufacture of superconducting (SC) magnets for an accelerator is mostly confined to large laboratories, where magnets are conceived and designed and where research and development (R&D) on small models and prototypes are carried out. It is only when it comes to series production for large projects that industry becomes involved. The lack of continuity between large projects, the need for special tools and the strong interlink between manufacture and design with a high degree of specific knowledge make it very difficult for industry to be able to maintain a team with the necessary expertise for accelerator magnet manufacturing. Actually, we can say that the number of companies capable of offering a real service to the laboratories in this domain has decreased with respect to 30 years ago, there probably being only three or four in the world at present.

## 2    A short history of the manufacturing of SC magnets

As soon as NbZr and $Nb_3Sn$ became available, see Ref. [1], researchers and engineers tried to make magnets for proton accelerators (for electrons the required field is typically 0.1–1 T so there is no need for superconductivity). Meanwhile, the newly discovered Nb–Ti became rapidly the option of preference; the use of SC magnets for synchrotrons, with fast ramping fields (0.1–0.5 T/s) was pursued, see for example the GESSIE collaboration [2]. However, all of these efforts have been carried out in research institutions. Unlike other types of magnets and applications, such as solenoids for research, power application, nuclear magnetic resonance (NMR) and magnetic resonance imaging (MRI), etc., accelerator magnet technology remained confined to research laboratories.

At the end of the 1970s CERN was able to boost the luminosity in the Intersecting Storage Rings (ISR), featuring 25 GeV proton energy per beam, by means of large-aperture SC quadrupoles,

---

of about 5 T peak field, placed in the low-β insertions [3]. In 1980 the eight, 1 m long, ISR quadrupoles were the first SC magnets in routine operation in an accelerator. Their cold masses were the first, small, SC magnet series industrially manufactured for accelerators.

Despite the success of the ISR quads, with the decision to develop the resistive Super Proton Synchrotron (SPS) and the Large Electron–Positron Collider (LEP) as the main accelerators for CERN for the 1980s and 1990s, the spotlight on the development of accelerator magnets passed to the United States, where two large new projects were 'racing' to be approved. At Brookhaven National Laboratory (BNL), Upton, NY, a design team was working on a 200+200 GeV proton–proton collider of 3.8 km based on dipole magnets operating at a 4 T field, later upgraded to 5 T. The R&D for this project was very important and many prototype magnets were actually built [4]. However, for various reasons the BNL project, called Isabelle and later the Colliding Beam Accelerator (CBA), was terminated in its infancy in 1983. Meanwhile, the Fermi National Accelerator Laboratory (known as Fermilab, Batavia, IL) in 1972 successfully commissioned its brand new 200 GeV synchrotron (based on resistive magnets, in a 6.9 km ring), which was later pushed to 500 GeV. The director, R. R. Wilson, launched a vigorous R&D program aimed at designing and manufacturing SC dipoles of 4–5 T in order to double the energy of the synchrotron [5]. His dream was realized in 1983, when the largest SC magnet system—composed of 774 dipoles rated for 4.2 T and 240 quadrupoles—was able to push the proton beam to a record energy of 517 GeV, to later reach almost 1000 GeV or 1 TeV, hence the name of Tevatron. The machine was later transformed into a collider, since the magnet bore (75 mm) was large enough, hosting two counter-circulating beams of protons and antiprotons. This gave access to ~2 TeV centre-of-mass energy: the Tevatron SC magnets kept Fermilab the leading high-energy physics laboratory until 2010, when the Large Hadron Collider (LHC) broke this record.

The Tevatron was the first very-large-scale demonstration of the possibility of superconductivity, with its 4.2 K cryogenics as a practical engineering system. From the design point of view the Tevatron SC magnets feature collared coils (CC) enclosed in the cryostat while the iron yoke stays at ambient temperature. This solution was possible because the forces were still manageable by the austenitic steel collars, but poses serious issues of field alignment and force interaction between coils and yoke. This solution was adopted because it shortens the cool-down time from 10–15 days to 2–3 days, thus reducing the machine downtime for interventions. In modern colliders the total number of collisions is the most important performance parameter, after collision energy. Because of this, reducing the downtime is as important as increasing the collision rate (luminosity). In the case of the Tevatron, given the numerous interventions that this pioneering machine required, the room-temperature yoke was probably a good choice. Anyway, one should remark that Tevatron magnets were built and assembled in the laboratory; industry was just providing mechanical and electrical components [6].

Although all construction was in the lab, given the large number of dipoles (774), the design and construction processes marked the beginning of industrially-oriented manufacturing with:

- use of 'simple' and relatively cheap Rutherford cable as conductor;
- use of collars to pre-stress the coils and contain the electromagnetic forces;
- use of laminations for collars, iron laminations and for manufacturing and assembly tooling; less well known, because it does not show up in the design, this last innovation was critical to achieve the requested tolerance at an acceptable cost.

The next large project was the Hadron Electron Ring Accelerator (HERA), which went into operation in 1991 at the DESY laboratory in Hamburg, Germany [7]. A single proton beam guided by SC magnets was made to collide against an electron beam. Of similar size and field to the Tevatron, the HERA dipoles featured an important difference: use of cold iron around collars that were made of aluminium alloy, to gain pre-stress during cool-down thanks to higher thermal contraction of aluminium than that of stainless steel. A cold iron yoke, following the Isabelle design, was the invariable choice of all main projects after the success of HERA. While the Tevatron magnets were manufactured in the laboratory, HERA was the first industrially built large project. About 420, 9 m long, dipoles and 220 quadrupoles, in addition to most of the 660 corrector magnets, were manufactured by European industry.

During the 1990s the Relativistic Heavy Ion Collider (RHIC) was built at BNL in the tunnel of the decommissioned Isabelle [8]. The field level was relatively modest for SC magnets: about 3.5 T. RHIC had a limited budget, so the challenge was to keep the cost low rather than achieve record-breaking characteristics. The 264 dipoles are each nearly 10 m long with an 80 mm bore, with cold iron construction. The most important characteristic is that the iron laminations, in addition to being the magnetic yoke for flux return, are used to restrain the electromagnetic forces, i.e. they are effectively the collars. The coils are surrounded by simple, thin collars, serving as spacers and for coil positioning and are made from a plastic material for cost saving. The iron is therefore very near to the collars, thus contributing noticeably to the field while the low field helps to keep saturation effects under control. The magnets are single-aperture so, to work as a collider, two complete, independent rings had to be built. The RHIC magnets have been built to print in Industry with a strong involvement from the laboratory.

In the US, after the termination of Isabelle and the start of the Tevatron, even before the design and construction of RHIC, a study was launched to design and build the 'definitive' machine: the Superconducting Super Collider (SSC), which had the goal of colliding protons against protons at 20 TeV per beam, an enormous leap for the discipline [9].

From 1983 until 1993, the year of its cancellation by the US Congress, the R&D for the 87 km long proton–proton collider marked the development of magnets and the first attempt of industrialization on a giant scale. To fill the 87 km tunnel more than 7700 dipoles, 15 m long and of 50 mm bore, were needed, and these were designed to reach 6.6 T at 4.4 K.

The design was based on austenitic steel free-standing collars surrounded by a cold iron yoke with no mechanical function. Among the breakthroughs carried out in the context of the SSC R&D have to be accounted: i) the increase of $J_c$ from 2000 to >2700 A/mm$^2$, 5–4.2 K, in conjunction with high-quality fine filaments of 5–6 μm; ii) the industrialization of high-quality cabling; iii) a new cable insulation based on full polyimide; and iv) a better way of collaring that was more suitable for industrial production. The SSC project was stopped in 1993 for various reasons, including the continuous increase of the cost and the presence of a less expensive, although more modest, program aimed at the same physics territory: the LHC.

R&D for the LHC at CERN started at the end of the 1980s [10] with serious funding allocated only after 1992, incorporating many features of various previous design studies and R&D (especially from the SSC). However, to take full advantage of the existing 26.7 km long LEP tunnel, CERN pushed Nb–Ti magnet technology to its extreme. Design innovations were: i) use of the two-in-one design first proposed by BNL [11]; the LHC was actually developed for the original 'twin' dipole variant, where the two channels are fully coupled both magnetically and mechanically; and ii) a coolant temperature of 1.9 K by using HeII to boost Nb–Ti performance and make use of the superior conductivity and heat transfer properties offered by superfluid helium.

The implications of the two-in-one layout on magnet design would go outside the scope of this work. However, its success is such that currently all magnets for future hadron colliders are designed with a dual bore, even for ion versus proton collisions, which is somewhat surprising given the difference in mass-to-charge ratio.

In Fig. 1 and Table 1 are reported a synoptic view of the cross-section of the main dipoles of the projects discussed above.

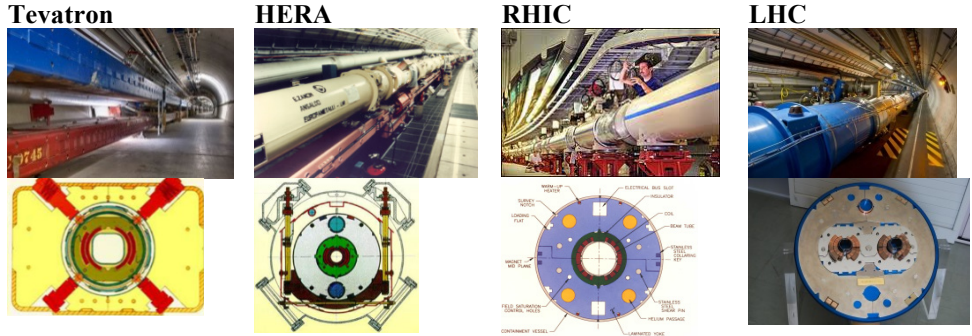

**Fig. 1:** Cross-section through the main SC dipoles of the SC colliders

**Table 1:** Features of the main SC dipoles of the SC colliders

|  |  | Tevatron | HERA | RHIC | LHC |
|---|---|---|---|---|---|
| Maximum beam energy | TeV | 0.98 + 0.98 | 0.82 (0.92) | 0.25 + 0.25<br>0.1/n + 0.1/n | 7 + 7<br>(4 + 4) |
| Dipole field | T | 4.3 | 4.65 (5.22) | 3.5 | 8.3 (4.8) |
| Aperture | mm | 76 | 75 | 80 | 56 |
| Magnetic length | m | 6.1 | 8.8 | 9.5 | 14.3 |
| Stored energy | MJ | 320 | 250 | 60 | 8000 |
| Weight (cold mass) | kg |  |  | 3,607 | 27,000 |
| Quantity | # | 774 | 422 | 264 | 1232 |
| Iron yoke |  | Warm | Cold | Cold | Cold |
| Operating temperature | K | 4.2 | 4.5 | 4.6 | 1.9 |
| Operating condition |  | He supercritical | He saturated | He forced flow | HeII at 1 bar |
| Structure |  | Single bore, warm iron, straight | Single bore, cold iron, straight | Single bore, cold iron, curved | Twin bore, cold iron, curved |
| Force retaining |  | Stainless steel collars | Al alloy collars | Yoke 'collars' | Stainless steel collars + shrink cylinder via yoke |
| Operating current | A |  | 5,027 (5,640) | 5,050 | 11,850 |

## 3 Cabling and insulation

Here we do not discuss the cable from the SC property point of view, but just as an element, and a critical one, of magnet construction. Regarding insulation, we will not discuss the electrical properties, rather we draw out the practical implications for coil technology.

### 3.1 Rutherford cable

The cable is a fundamental unit of the construction. A Rutherford cable is composed of twisted strands, see Fig. 2. Cabling can in part add or remove the strand twisting, which contributes to the mechanical stability of the cable. In Fig. 3 a cabling machine used for LHC cable production is shown. The most important characteristics of the cable from the point of view of the winding are the dimensions and mechanical stability under tension and bending.

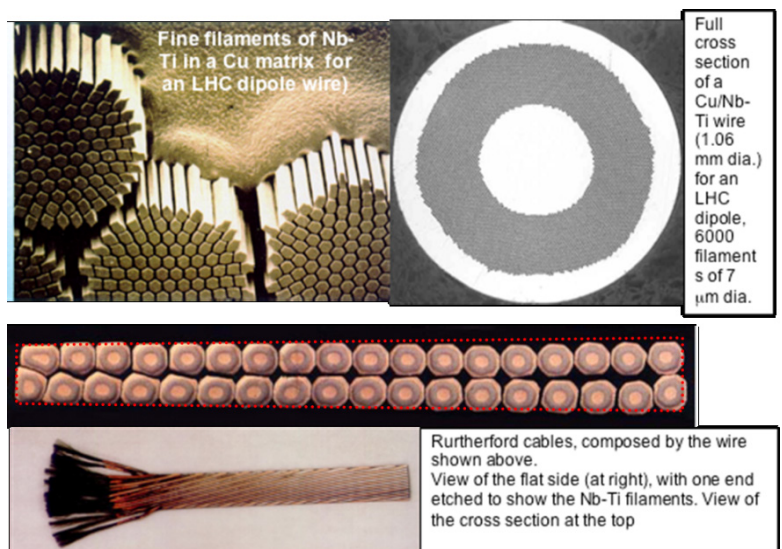

**Fig. 2:** Composition of the LHC dipole Rutherford cable

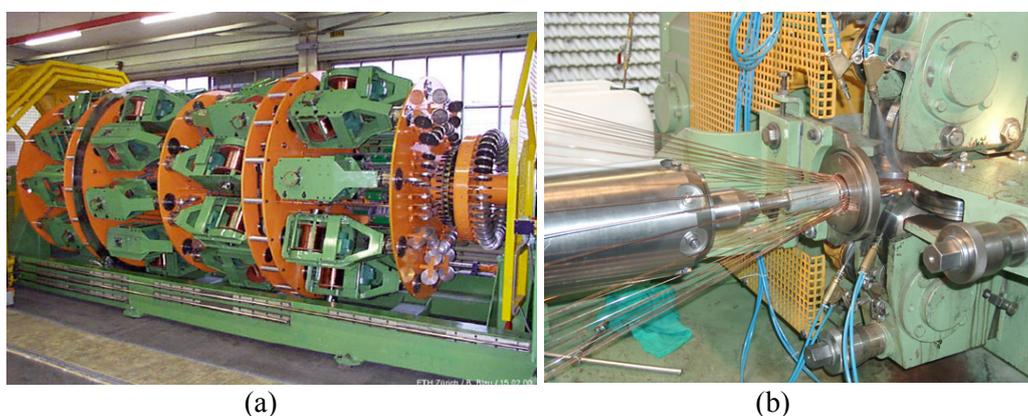

(a)            (b)

**Fig. 3:** Cabling machine for the LHC Rutherford cable: (a) view of the large rotating wheels carrying the strand spools; (b) view of the 'turk head' where the cable is formed.

### 3.1.1 Dimensions

The collar concept relies on enclosing the coil package inside a cavity whose dimensions are given by the collars. The typical accuracy required by magnet designers for the coil package is about 50 µm or better. Since the number of turns in a coil (see Fig. 4) is of the order of 20–40, in principle the accuracy required of a single unit, the cable, would be 1 µm. Actually, the accuracy of a strand is of that order and the accuracy of the cable cannot be as good. However, in series production what matters is not, to a certain extent, the absolute value. The average thickness of a cable is a value that, within limits, can be corrected or compensated for with a systematic action (for example, by adjusting the size of the collars or by shims between the collars and the coil): therefore, what is important is that the variation is kept within very tight limits. In Fig. 5 the actual variation in one of the cable production lines for the LHC is shown. It shows an amazing uniformity of better than ±2 µm. However, this is not the thickness 'seen' by the coil manufacturers. The values in Fig. 5 are measured just at the end of the cabling line, before any manipulation, before the thermal treatment for inter-strand resistance control and before the cable is unwound from the spool and then wound onto the coil. Being a composite and not a solid piece, the cable size depends on the conditions of measurement in a very subtle way. So in the coil winding process it is important to have a means of accepting a different average value (also 10–20 µm, which translates to 0.5 mm of the total thickness) while the handling, tooling and processing must ensure that the transverse pressure and longitudinal tension be always uniform. Only in this case can we profit from the uniformity shown in Fig. 5 for manufacturing coils that have dimensions as uniform as possible. As a last remark it should be remembered that the Rutherford cable must always be measured at the reference transverse pressure, the one used for coil-curing (Nb–Ti) or treatment (Nb$_3$Sn), and in the correct conditions of side confinement, in order to have consistent data.

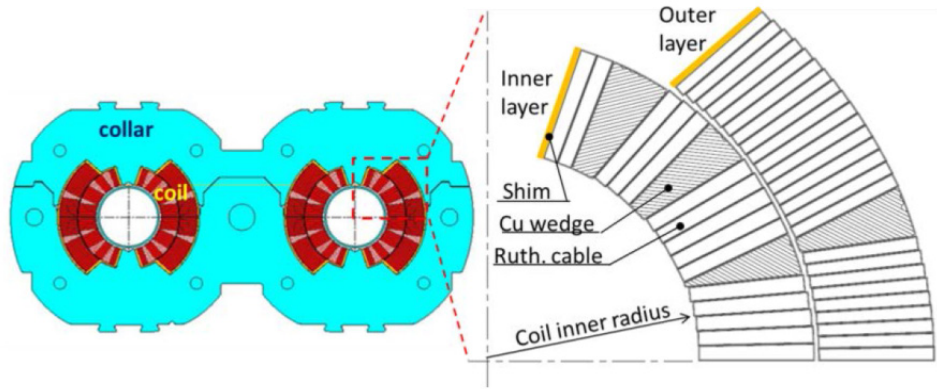

**Fig. 4:** LHC twin dipole CC with an expanded view of a quarter of a SC coil

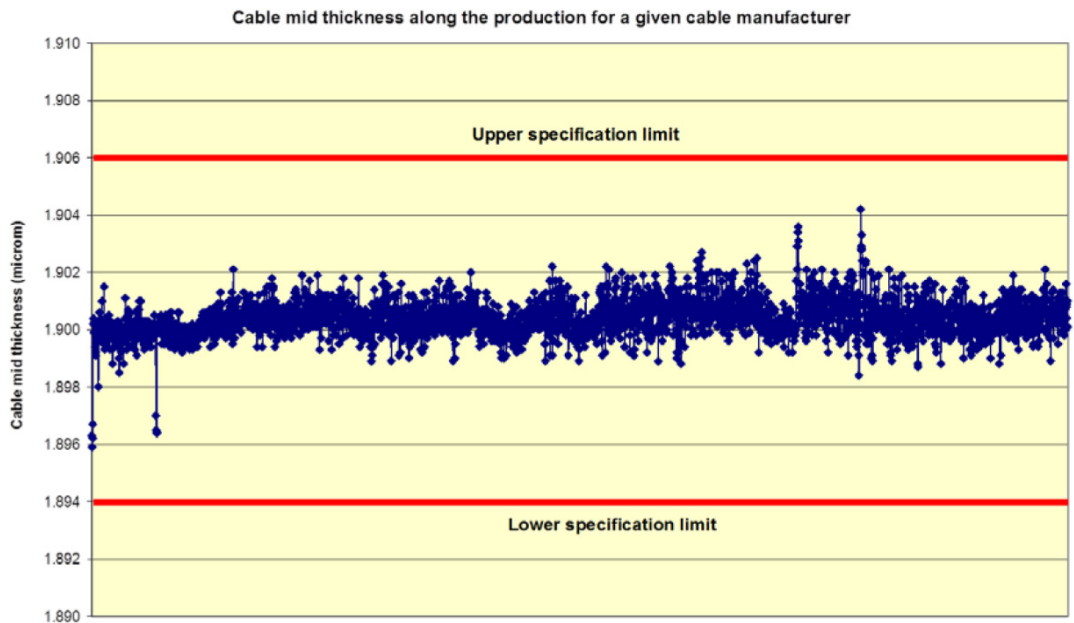

**Fig. 5:** Cable thickness measurements from a production line for LHC dipole inner cables. More than 1000 km of cable are represented here.

The mechanical stability of the Rutherford cable is also very important. A Rutherford cable is a loose ensemble of strands, plastically deformed at each folding (but with some spring-back if a thermal treatment is not applied), with no glue between the strands. A Rutherford cable is stable if it is contained by means of suitable tooling or if it is tensioned with the right force: too small a force will leave the strands free from each other and a too high tension will drive the cable near to an instability point and the cable can suddenly buckle into a round cord. This second case is a disaster that is usually detected during winding: depending on the degree of buckling and the length, and upon the skill of the operators, the cable may be recovered. The first type of instability is subtler: during winding, a strand can bulge out, especially in the place where winding tension changes, such as in the bends. If the defect is identified during winding it can be repaired, by experienced operators, by unwinding and then pushing back the strand(s) popping out and restarting with more care: it takes time and increases costs. However, the strands are below the insulation when the coil is wound, and the defect may well pass undetected. In such a case the cable may be degraded, more or less severely, during curing or thermal reaction. With Nb–Ti the degradation in terms of critical current is not too severe [12]; however, the real danger is possible damage to the insulation. Cable mechanical stability must be tested *a priori* and specified at the bidding time, as was done for the LHC. We conclude this section on cable stability by remarking that the Rutherford cable has a rotation, or transposition sense: it is built with left-hand or right-hand torsion, see Fig. 6. It is easy to realize that if the coil is wound with the same rotation sense of the cable torsion, the cable itself tends to increase its stability, while if it is wound in the opposite way its stability is decreased. If each layer is wound independently this is not a problem: winding sense and cable torsion sense can be adjusted. However, if the coil is built with the double pancake technique, i.e. without a joint in the layer jump, one layer is wound in a favourable sense and one in a less-favourable sense. For the LHC this was an issue at the beginning of prototyping when the cable was not yet optimized. During manufacturing it remained a concern only for the main quadrupoles, wound in a double pancake: in general, the series production LHC cable was so well optimized and mechanically stable that this did not cause much trouble, with the exception of the cable from just one manufacturer.

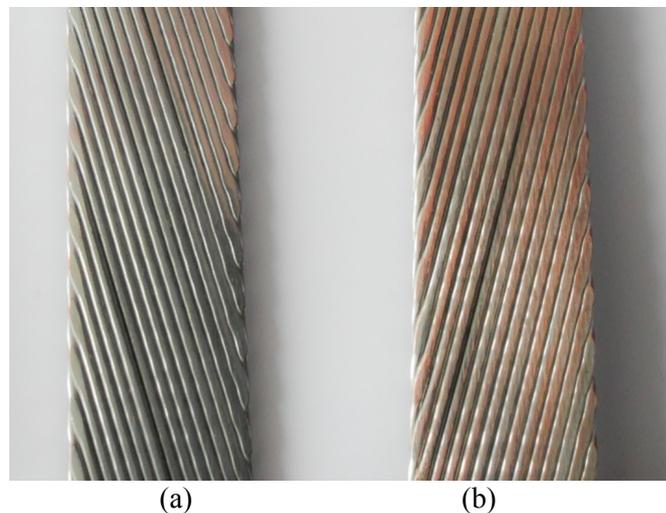

(a)                    (b)

**Fig. 6:** (a) Left-hand and (b) right-hand torsion cables for a LHC dipole. Cable (a) is from LHC production, while (b) dates from 1996 and was used in the first LHC 15 m long prototype (CERN–INFN collaboration). At that time cables had a long transposition pitch and were not oxidized.

The cable is by far the most precious component of a SC magnet: the cable makes about 20-30% of the cost of a Nb–Ti magnet; this cable cost fraction can reach 50-60% of the total cost for $Nb_3Sn$ magnets. Therefore, the first manufacturing issue is avoiding the waste of cable. A good practice is using dummy copper cable (better if alloyed to have better mechanical properties) for debugging and then tooling up for the production process. Another, even more useful, practice is the use of super-dummy cable (low-value superconductor, of degraded SC performance but of near-to-final mechanical properties) for fine adjustments to the whole coil manufacturing process. In many cases a super-dummy coil and coil assembly may also be useful for testing the entire cold mass assembly and even when debugging the testing station and procedures.

### 3.2 Insulation

For insulation, dielectric strength put aside, the problems are of two types. The first one concerns the ability of reaching the desired geometrical tolerance. Each cable is covered with insulation of a thickness of 70–150 μm. If we take an average of 2 × 125 μm for an LHC dipole (each cable has two faces) and compare it with the cable thickness (see Fig. 5) we see that it is about 10% of the total thickness. However, while the cable can be accurate at a level of ~0.1%, manufacturers of insulation sell it with a variability of 5% or worse. So the uncertain insulation thickness may dominate the geometry of the coil. Measures can be taken, by working with the supplier to better qualify the process, both by intrinsic improvement of the uniformity and by sorting production into batches of different average thickness and small variance. It is important anyway to measure the thickness of stack of many insulation layers and, finally, to measure the thickness of insulated cable stacks (typically 10 cables) under transverse pressure and conditions similar to those experienced during curing and/or thermal treatment. During curing of polyimide or epoxy pre-impregnated insulated coils, insulation flows at high temperatures, partially filling the interstices between adjacent strands, thus impacting both on cooling and on the final effective dimensions of the coils.

A phenomenon that might be of considerable importance for large projects is insulation creep. Creep is usually important only at high temperatures; however, if a magnet is stored for a few years in improper conditions, as happened to the main LHC dipoles because of delays to the test station and later due to tunnel availability for installation, creep might play a role.

Another practical issue for insulation based on polyimide, such as the LHC main magnets, is the careful control of the polyimide glue on the external layer. Glue, which is activated by curing at about 180°C, is necessary to handle the coil until collaring takes place. The tape on the glue is wound with a gap to create the very thin channels (2 mm wide, 0.02 mm thick) in which superfluid helium can flow to cool the winding. The correct curing of the glue happens in a relatively narrow window of temperature and humidity. This last factor is usually neglected but it can have a detrimental effect. In particular, low humidity can, surprisingly, disable the curing effect. A careful check of the uniformity of the quality for the glue was important during LHC construction. Due to some alarming results, sampling of the glue quality, done via a peel-off test, was reinforced, which led to the detection of a serious degradation of the gluing factor with time. Certain deliveries had to be refused, and a stop to coil production due to a lack of adequate insulation was avoided with a very small margin.

To conclude this section about insulation, we would like to pass on our experience accumulated in many large and small projects.

  i. One must be generous with insulation, inter-turn, inter-layer and to ground: an insulation defect, in the presence of high $J_c$ and large amount of stored energy, can suddenly and unexpectedly damage the magnet, without any precursor. Usually, a mechanical default and/or quench bad behaviour degrade the performance of the magnet, and only rarely do they prevent the magnet from working at some useful level. However, an electric fault may easily make the magnet completely unusable.
  ii. Insulation must be the responsibility of the magnet manufacturer, avoiding additional interfaces, outsourcing and dilution of responsibility. In the LHC project, CERN provided all insulation material. However, the magnet manufacturer was responsible for a careful inspection of the insulation process (done by subcontractors), as well as a thorough check of the conformity of the insulation to the specifications and for its suitability for the actual construction and operating condition of the magnet.

# 4     Nb–Ti technology and the LHC experience

Nb–Ti technology owes a large fraction of its success to the ductility of the Nb–Ti cable: the single wire behaves like a copper-reinforced wire and can survive mishandling or process mistakes. Really large errors and huge mishaps are needed to produce damage to a Nb–Ti wire or cable.

The construction of an accelerator magnet can be subdivided in two main parts: CC and cold mass completion (CMC). Here we review the various steps for each part.

## 4.1 Collared coil

### 4.1.1 Coil winding

The single coil unit is a pole coil that can be composed of a single layer or double layers. In the latter case the coil can be fabricated with the double pancake technique if a single unit length is to be employed for the whole coil, or by joining two separately wound coil layers. In this case, also called a 'false double pancake', a different cable can be employed allowing current density grading. In practice, accelerator coils are composed of one or two layers. There are a few cases of four-layer coils, which are usually composed of two double pancakes. Coil winding must be very accurate (the tooling must ensure about 20 μm accuracy over 15 m, which can be done at an affordable cost only by assembling laminated components; see Section 2). A double pancake is very suitable when the azimuthal extension of the two layers is similar, otherwise the insulation curing (see below) may become difficult. The most critical issues are as follows.

  i. Avoid damaging insulation with sharp-edged metal (but plastics may also be dangerous).

 ii. Ensure correct, uniform tension of the insulated cable both in the straight parts, in the ends (submitted to a 3D bend) and when tension is transferred from the spooler to the retaining device. These are usually pneumatically activated pistons, pushing on the straight part after or before the end bend is engaged.

iii. Avoiding strand pop-out (see the section above on the Rutherford cable).

 iv. Make good end parts. The saddle-shape ends needed for our magnets are actually one, probably the main, source of quench. Although optimized codes to design them are available, making a good end is still an art requiring more than one iteration.

  v. Formation of the transition between layers, usually called layer jump.

### 4.1.2 Coil curing

This is a moderate thermal treatment (150–190°C) under pressure to cure, or polymerize, the glue that is put on the external layer of the cable insulation. By gluing the layer in the right position, the coil acquires a consistency and a shape to withstand all manipulations until the coil is collared. Polymerization is the second critical step, not only because of the sensibility of the component to its quality and to various environment conditions, but also because the accuracy of the polymerization tooling determines the precision of the final coil. For this reason the polymerization pressure is very high, not far from the pressure needed for collaring, and much care is taken in the design of the curing mould (or cradle). For LHC coils pre-polymerization happens under transverse (azimuthal) pressure up to 100 MPa at 100–135°C and then final curing under 80 MPa transverse pressure at 190–193°C. The pressure corresponds to a vertical pressure of 2.5 MN/m (single coil). In the case of a two-layer coil (or two double-pancake coils, or a four-layer one) one can choose between the following options.

  i. Winding the outer layer on the top of the inner one and then polymerize the complete coil. This is the only possibility for a real double pancake with no electrical junction between the two layers.

 ii. Polymerizing the inner layer right after winding, then winding the outer layer onto the top of the inner one, finally polymerizing the whole coil. In this way the inner layer is polymerized twice: however, this ensures better control of the accuracy of the absolute value of the radius and the variation of the second layer. Also, this procedure gives more flexibility to production lines for a large quantity, allowing better tooling specialization. This procedure also allows the assembly of quench heaters on the top of the inner layer if necessary. However, it requires soldering the cables of the two layers, always a delicate operation liable to weaken the electrical integrity of the coil, becoming a possible source of heating, reducing the stability margin (in the case of a bad joint). In any case, the electrical joint is a singular point in the mechanics of the coil assembly.

### 4.1.3 Pole assembly

This is the work required to make the coils ready for collaring: installation of coil instrumentation; assembly of the outer layer on the top of the inner one, if separately polymerized; soldering of the cable for connecting the inner to the outer layer (if not a double pancake); and insertion of the interlayer insulation, which is usually a critical component that must ensure electrical insulation while allowing a generous flow of coolant for the required heat removal from the outer coil. The following step is the assembly of a critical component: the quench protection heaters. They are usually placed on the outer side of the external layer; in the LHC main dipoles there are two of them, for redundancy, which means that the coil azimuthal extension of a single quench heater is less than half. Quench heaters, a metallic strip encapsulated between two polyimide layers, must be thin for fast heat diffusion into the superconductor, however they must also be robust enough to withstand a pressure of 150 MPa and thermal contraction, which are contradictory requirements that require a compromise to be found. Various layers of ground insulation, typically in polyimide thin foils, are then placed and a coil protection sheet is eventually laid onto the ground insulation to make sure that the high-pressure during collaring is uniformly distributed, avoiding peak stress concentrations.

### 4.1.4 Collaring

Collaring is a critical operation because, in most types of design, it encases the coils in their force-retaining structure and also gives the final shape to the coils, thus playing a big role in the main performance indicators of a SC magnet: quench behaviour and field quality. The collaring operation is carried out under powerful presses, see Fig. 7, and in the case of the LHC about 300 MN are applied to the 15 m long collaring to obtain the target pre-stress [13]. All components are manufactured to an accuracy level of 20–30 μm, and assembly aims at 50 μm final accuracy. The accuracy of the collar profiles through production of the LHC dipoles is shown in Fig. 8: about 12 million collar pieces have been manufactured in special austenitic steel through a fine blanking technique. In order to preserve this accuracy of components in the finished magnet, the hardened steel beam that transmits the force from the press pneumatic pistons to the collar–coil package is a single piece, 300 mm thick, 700 mm wide, machined along the 15 m length to an accuracy of 20 μm.

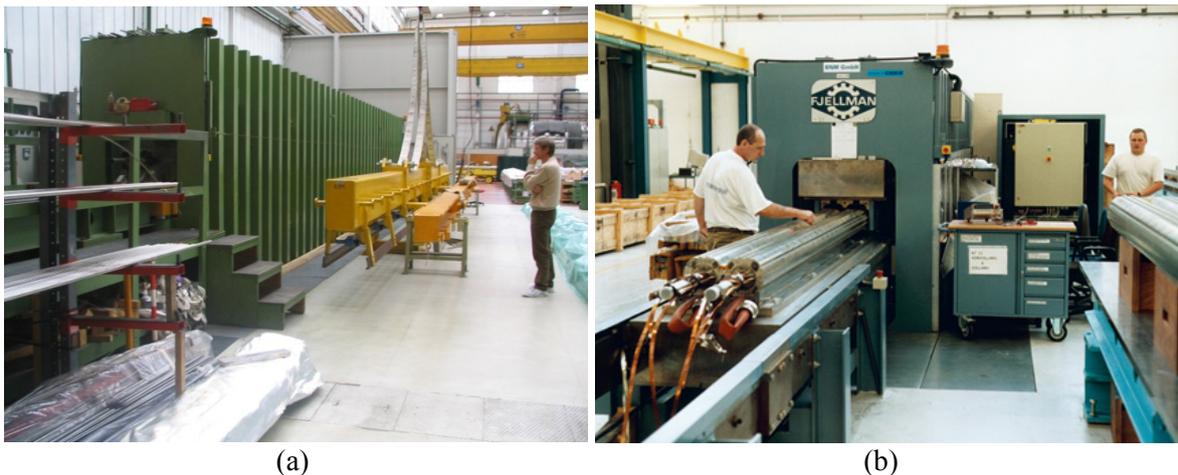

(a) (b)

**Fig. 7:** (a) A 15 m long collaring press for LHC dipoles, capable of 2100 tonnes/m (~20 MN/m) force; (b) insertion of a CC assembly into the press for performing the collaring operation.

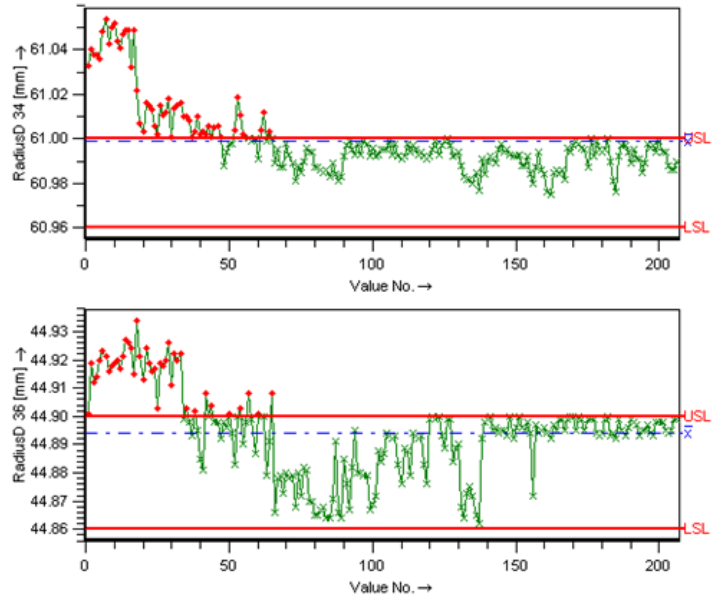

**Fig. 8:** Evolution of two important dimensions of the LHC dipole collars through production (Value No. indicates the batch number), showing the ability to stay, after an initial correction, in the desired range of ±20 μm.

The coil must be very accurate too, and its radial thickness and azimuthal extension must be near to the absolute target value, to avoid being damaged by the collar closing. But even the most precise coils cannot become their final size until pressure is applied. In other words, coils change size during collaring, passing from zero to peak stress of 150 MPa (see [13, Fig. 19]). The movement of the conductors can range from one to several millimetres, in the case of coils with spongy insulation. Collaring must be accomplished slowly, massaging the coils through cycles: during each cycle the press is released after locking rods or locking keys are inserted when the dimensions are near to those of the final value. Collaring is best accomplished by carefully choosing the point of pressure application, in a way to minimize the stress lost after releasing the force by the press; and the job of keeping pre-stress is passed over to the locking rods or the locking keys. Further discussion depends on the details of the coils, the collars' shape and the press.

A final word on collaring is about the style of manufacture. For a dipole a horizontal axis press naturally pushes in a vertical direction. However, this choice is less evident for quadrupole. To respect the four-fold symmetry a solution is to have a press with a vertical axis, and to apply pressure along two perpendicular directions at the same time. Devised at the Commissariat à l'Energie Atomique (CEA), in France, many years ago, this system has been applied to most quadrupoles produced so far, with excellent results. However, it inconveniently requires positioning the coils in a vertical direction; therefore, the height available above the floor must be two times the length of the magnets, to allow the coil to be inserted into the press. Quadrupoles are usually shorter than dipoles: the LHC main quadrupoles are 3.5 m long, for example, and this is not too difficult. It would be complicated for the next generation of IR quadrupoles for the LHC where low-β quads (if in Nb–Ti) more than 10 m long would be required, unless the magnet is further subdivided into subunits. For such a size, a horizontal press like that used for a dipole would be better to avoid enormous vertical infrastructure. Of course this technique requires that the uniaxial (vertical) force is transmitted to two orthogonal axes at 45% from the vertical. The system, first proposed for the low-β of the SSC [14], and also for the first design of the low-β quad for the LHC upgrade [15], has to be carefully tuned in detail to avoid a rise in unwanted spurious components.

An interesting alternative path is the use of the iron yoke as a collar. For dipoles this is a possibility for low-field magnets, where iron saturation is not severe: iron must be radially very near to the coils to be effective, and saturation may degrade field quality. The use of such system for quadrupoles has been carried out for the MQXA magnets, a type of LHC low-β quadrupoles, by the KEK, the High Energy Accelerator Research Organization, in Japan [16]. As with a horizontal press, the problem is that the main force is uniaxial: a system to project it on two perpendicular axes 45° from the vertical is necessary. Again, the results have been very good, but not without some corrections for the rise of unwanted harmonics.

*4.1.6 Coil enclosure*

The CC package is quite robust and can be easily handled. Many measurements are carried out on this package (see below), which then has to be completed with the insertion of the cold bore tube (CBT), which is the beam pipe. Since aperture is one of the most precious assets of an accelerator magnet, the room left between the CBT and coils is usually minimal. For the LHC dipoles, which are 15 m long, the solution was to assemble the coils before collaring around the CBT, so it remains trapped inside the coils after collaring. The disadvantage of this solution is that an intervention on the CBT requires a de-collaring of the magnet, which is a big operation and not without a risk of damaging the coils themselves. In the case of the LHC quadrupole the smaller coil length against that of the dipole (3.3 m against 14.3 m), and the slightly larger margin between coils and CBT (gained by reducing the insulation of the CBT, which was possible because of the lower quench voltage of quadrupoles than of dipoles) has allowed insertion of the CBT inside the coils after collaring. The operation could be repeated for the very few quadrupoles for which it was necessary, but not without damaging the CBT insulation, which had to be replaced (insulation is so hard to remove from the CBT that its re-use is not convenient).

The collar package is then finished by closing the coil ends with a plate: this plate is the interface between the coil ends on which the longitudinal force is exerted (see Fig. 9) and the blocks or screws that oppose this force by loading the end flange of the cold mass. This CC flange also has to allow an exit for the instrumentation wire and the quench heaters' wiring.

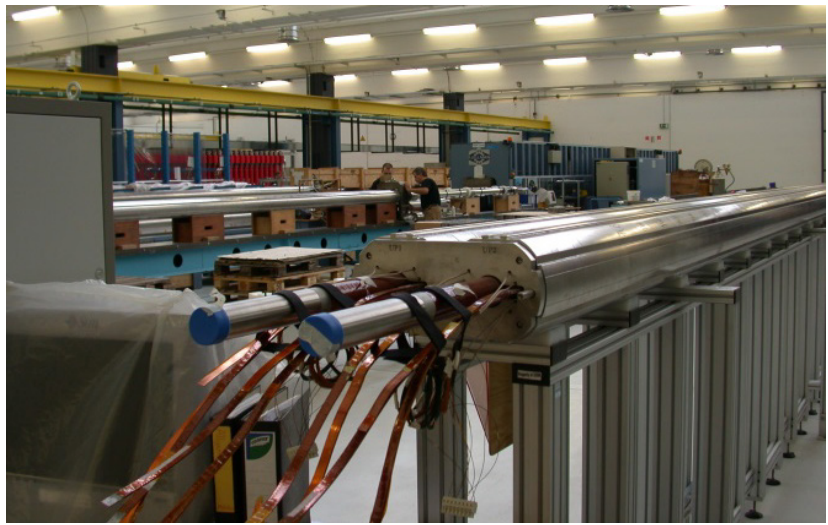

**Fig. 9**: Picture of an LHC dipole CC, finished with the end plate at its extremity

## 4.2  Cold mass completion

*4.2.1 Yoke assembly*

The first operation is to surround the CC with their magnetic circuit. Accuracy of the iron lamination composing the yoke is typically 50 μm, which also calls for a fine blanking manufacturing technique. This is very important, especially when for cost-saving reasons the thickness is rather large: in the LHC dipoles it is almost 6 mm (in resistive magnets the laminations are typically 1–1.5 mm). If some parts of the yoke contribute to the mechanical structure (see, for example, the iron insert in the LHC dipole cross-section, Fig. 10), their accuracy is similar to that of one of the collars, ~20–50 μm. Iron laminations are covered with paint, or a passivation (phosphatation) or oxidation layer, in order to avoid eddy current build-up and rusting: however, rusting typically occurs on the lamination edge, on the surface exposed to the cut. An important feature of the yoke is that the filling factor must be under control: in relative terms for the field homogeneity between magnets, since the iron yoke contributes approximately 15–20% to the total field, and in absolute terms because it affects the spring-back after magnet curvature (see below).

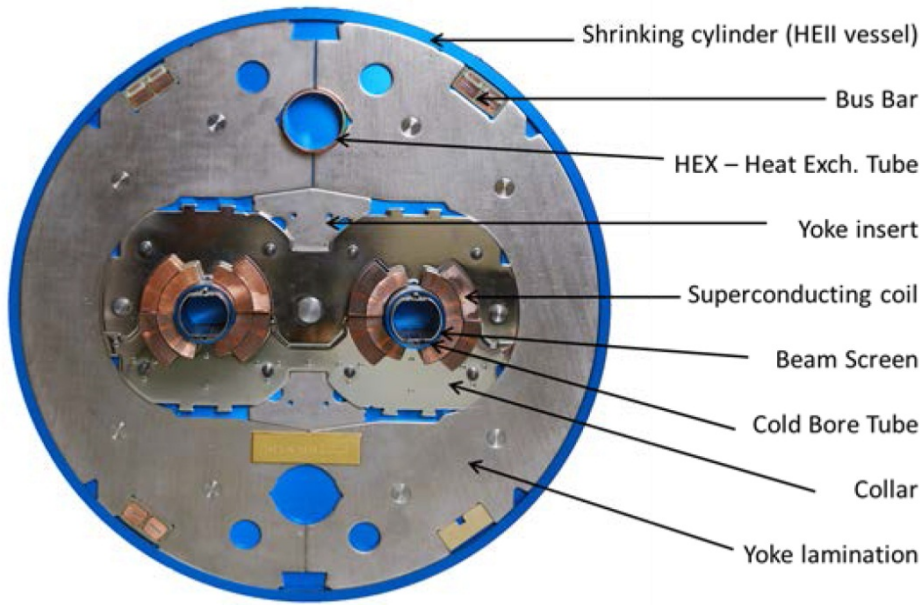

**Fig. 10:** Cross-section through an LHC dipole (magnetic circuit or cold mass)

### *4.2.2 Restraining cylinder/helium vessel*

The external cylinder has the primary function of being the lateral surface of the cylinder that constitutes the helium vessel, as in the case of the HERA dipoles. However, different functions can be assigned to the external cylinder by the magnet designer. For example, in the case of the LHC main quadrupoles it works as an inertia cylinder: it ensures straightness along the same axis of the main quadrupoles and the corrector package (octupoles, trimming quadrupoles, sextupoles, and orbit correctors' dipoles). However, it is not required to contribute to coil pre-stress. For this reason it is a 16 mm thick 316LN steel single-piece cylinder, with precise pins for magnet precise positioning. Such a choice is naturally coupled with vertical assembly, requiring large infrastructure (see Fig. 11).

In the case of the LHC main dipole, the cylinder is required to assist the collars by exerting an additional pre-stress on the coils, via the yoke. Such an assembly is also called 'line fitting' because the relevant surfaces must be very precise and fit (see Fig. 12). The pre-stress is given via shrinkage provided by welding two half-cylinders that have an azimuthal length shorter than the azimuthal length of the outer yoke. A discussion of this point and of its kinematics is beyond the scope of this paper. It is sufficient to say that, to ensure the required uniform shrinkage in the presence of a geometry that cannot be as precise as desired with 16 m long open half-shells, CERN has developed a new welding methodology by adapting the surface tension transfer (STT) technique to austenitic steel and to synchronous double-side topology. STT is an advanced fast method for oil pipes, and it has been developed for carbon iron with one single-side welding. This has been a great effort because a fully robotized system integrated into the huge presses (220 MN over 15 m with an accuracy of 50 μm for a beam cross-section of 600 mm × 300 mm but free frame of 1.5 m × 1 m) had to be developed and industrially commissioned where the cold mass construction was taking place. The welding press and welding operation is shown in Fig. 13.

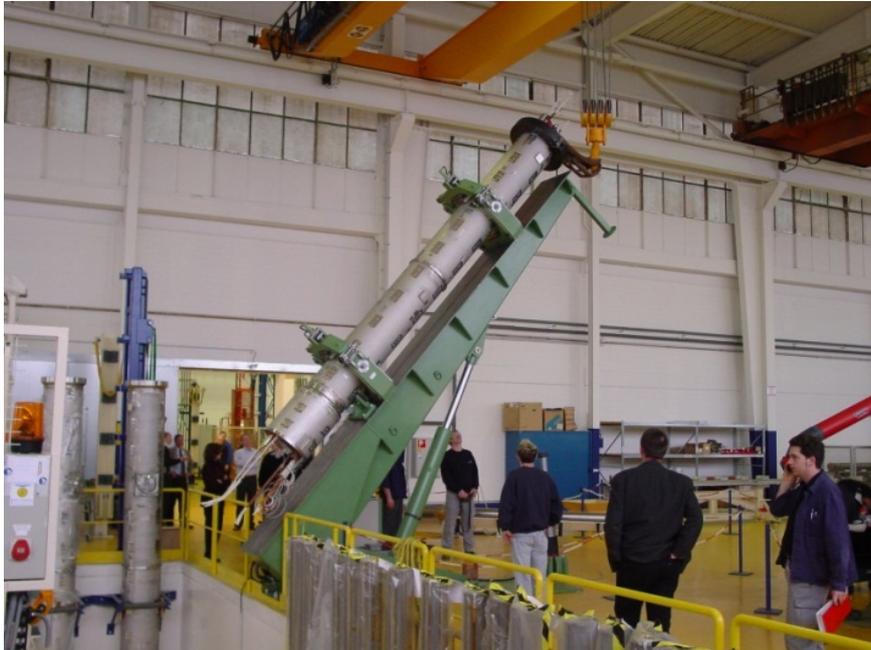

**Fig. 11:** System to turn the LHC Quadrupole cold mass vertically (courtesy of Accel, Germany)

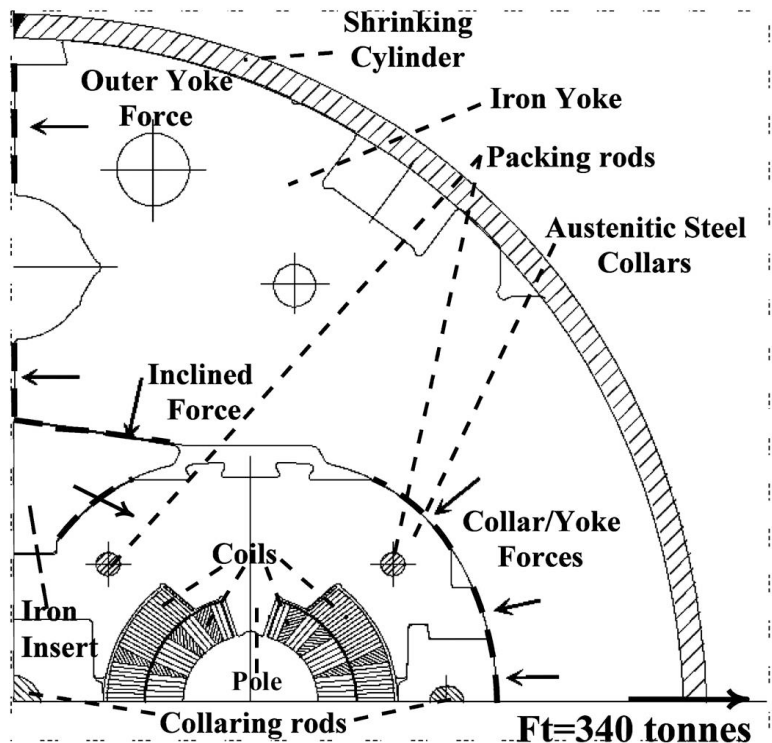

**Fig. 12:** Quadrant cross-section through an LHC dipole, with the precise (25–50 μm) fitting surfaces indicated with dashed lines.

This step of the longitudinal shrinking cylinder welding also serves to curve the magnet. A sagitta of 9 mm is generated on the 15 m long, 600 mm diameter cylinder. The magnets are welded with an over-bend sagitta of about 15 mm; then the spring-back due to welding shrinkage and to the spring effect of the iron laminations (which are more compressed on the inner side) reshape the magnet at, or near to, the target curvature. Actually it was almost impossible, despite all attempts, to attain the severe ±1 mm sagittal tolerance and even more difficult to achieve the ±0.3 mm tolerance required at the extremities. However, an ingenious way of assembling the magnet inside the cryostat and a sorting during installation has allowed achievement of and even exceeding the target alignment tolerances, with the result that today the LHC machine is better aligned than expected, with beneficial consequence for the beam dynamics.

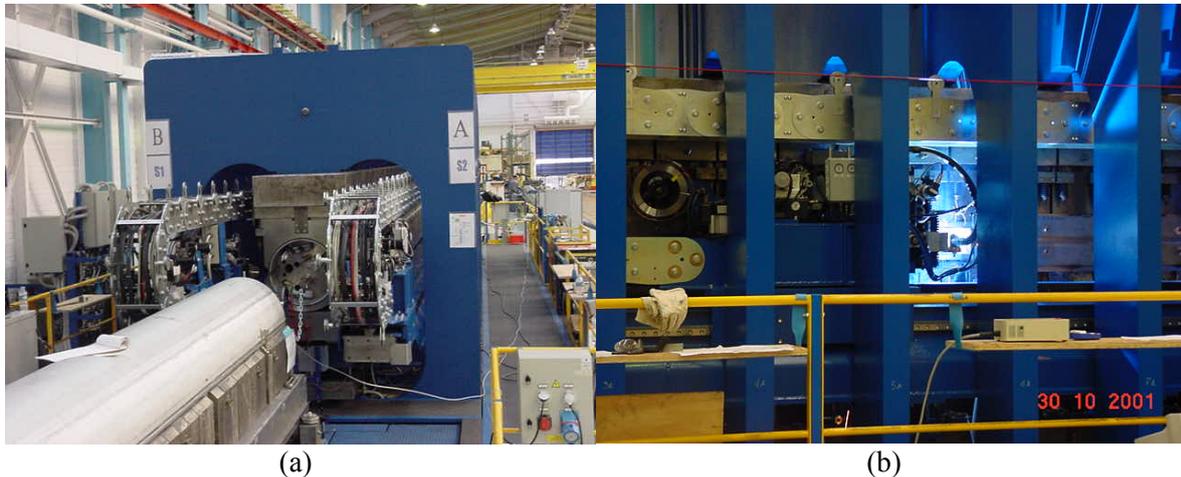

(a)                                                                 (b)

**Fig. 13:** Automated dipole welding in the large press during series production: (a) front view of the welding press, with one dipole under the press enclosed in the cradle mould during welding, and the next dipole waiting to be inserted; (b) side view of the robotized welding operation.

### 4.2.3   Connections and bus bars

The various coils need to be connected to form the circuit and then connected to the by-pass line for protection (bus bars and diode), see Fig. 14. The connections must be stabilized and secured against electromagnetic forces. Lack of space makes it difficult to integrate the various components; and even standard processes and operations, such as TIG welding, may become very difficult, requiring special procedures and very skilled personnel. In addition, stability and reliability of connections that have to carry 13 kA under various conditions (including a 50 mm elongation when passing from 300 K to 2 K), and submitted to a violent heating and temperature gradient in the case of quench, should not be underestimated. In the LHC dipole this part is less than perfect and it may eventually be a limitation of the whole accelerator.

The various connections to the diode are not well engineered, and the surface resistance under certain conditions may increase outside the control limits; the diode package and the various copper buses have an electrical insulation that is far from robust. Naked surfaces are dangerously close to the enclosing steel box (ground), making this system vulnerable to shorts caused by metal chips. Indeed, the quantity of steel chips and burrs that have been trapped in the magnet is much greater than expected. This is certainly an unexpected drawback of having a thick welded cylinder in the presence of a diode circuit that is not well protected and in the lowest point of the magnet. During the first commissioning of the LHC, following the strong He flushing, necessary for the cool-down, we experienced a few electrical short-circuits. We expect that occasionally, after future cold-down–warm-up cycles or after the strong mass flow induced by a quench, the LHC dipoles may again occasionally experience this type of short.

### 4.2.4   Corrector magnets

It is normal to assemble a good fraction of the correcting circuit directly on the main dipoles, to have a chance of locally correct systematic effects (either geometric or persistent currents). Since these are small magnets, they are assembled around the beam tube and are called 'spool pieces' to distinguish them from other correctors or higher-order magnets that are placed in the corrector package that is usually placed together with the main quadrupoles. In the LHC all dipoles have sextupole correctors (providing an integrated sextupole strength equal to that given by the whole of the lattice sextupoles) on one extremity of the cold mass; in addition, each second dipole features, on the other extremity, a small octupole–decapole correction package.

The main difficulty is placing the magnetic axis of the spool piece at the right position, with a tolerance that for the LHC is ±0.3 mm. The connections are not difficult but their numbers are enormous—about 60,000 inside the cold masses: since some families of corrector magnets are series-powered over the whole sector (⅛ of the LHC), one single bad connection kills ⅛ of the correcting strength, which is not acceptable for LHC operation.

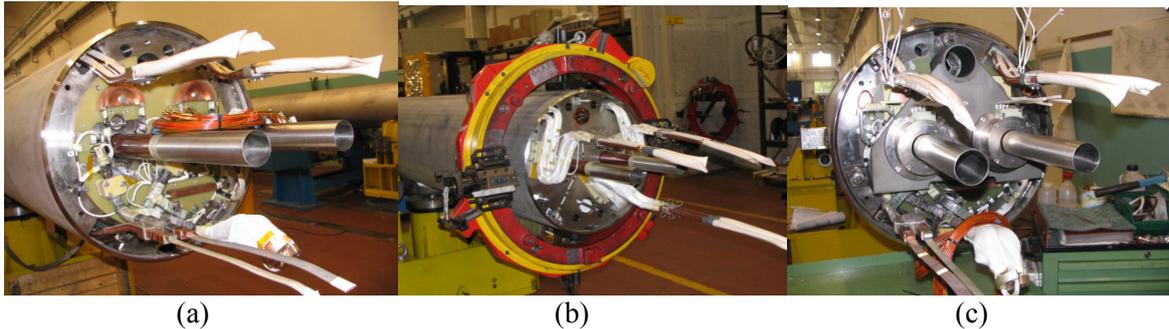

(a) (b) (c)

**Fig. 14:** Extremities of an LHC dipole (before closing the cool mass), showing: (a) the connections between various coils; (b) the lyra-shaped bus on the other extremity of the dipole; (c) the corrector magnet assembly.

### 4.2.5 3D part

The finishing of the end part is technically less interesting from the point of view of the magnet design; it is, however, critical for the success of the magnet's operation. We review here the main operations.

   i. Insertion of the Heat EXchange (HEX) tube, the complex copper-stainless steel vacuum brazed tube where the superfluid helium (HeII) is actually formed and that takes the thermal load of the magnets at 1.9 K.

  ii. Force retainer closing flange. This is a very thick flange that is placed against the coil closing flange, to support the longitudinal force. This stainless steel flange is 50 mm thick in the LHC since it has to support a force of the order of ~10 tonnes without bending. It is pre-loaded against the end covers (see below) with adjustable screws.

 iii. End covers and weldings. LHC dipoles and quadrupoles are closed longitudinally by end covers, which are almost all dish-shaped: welded to the external cylinder they form the HeII enclosure. In the LHC there is more than 40 km of thick stainless steel welding that is HeII-tight at 20 bar (pressure peak during quench). In addition, there are some 40,000 welds of various bellows of various sizes and other delicate welds, such as the non-fully penetrating fillet welding on the beam tube that separates the superfluid helium from the beam ultra-high vacuum and from the cryostat insulation vacuum. All must be HeII-tight under 20 bar operation pressure (quench) and many have severe alignment and dimensional tolerances. To have success a strict campaign of qualification of the process and of the people executing the work was put in place, using advanced technology. It is interesting to note the special development for the end covers of the dipoles. The end covers were produced with powder sintering technology, the first application on 40 mm thick 316LN of 600 mm diameter, which has won a special prize for CERN and the Finnish company METSO.

iv. Cabling and instrumentation. In the LHC magnet series construction, the instrumentation was reduced to a minimum, in order to reduce risk and cost; a decision that we partly regretted at a later stage, during magnet testing and accelerator commissioning. However, numerous wires for quench detection (redundant), for temperature measurement, for quench heater powering, and for diode control had to be properly installed using a minimal space. The exit port of the LHC instrumentation is not really satisfactorily arranged, also because of the tight space, and numerous interventions were required to fix problems during test and commissioning.

In the CMC, and especially in the 3D part, the complexity rather than the technical difficulty plays a major role. It is the realm of integration rather than technology, and somehow appears less challenging. For this reason it is often engineered too late, when good solutions are no longer possible. If there is an invariable lesson coming from all projects, it is that the integration studies should have started at an earlier stage.

## 5    LHC construction experience

### 5.1    Quality assurance and quality control

A quality assurance and quality control (QA/QC) plan can only follow a compiled list of the manufacturing steps, an analysis of what is really important to measure, a decision about what are the acceptable tolerances, and a clear definition of what to do when the measured values are outside the target values. All of this can be done only if the design is properly elucidated in all aspects. The technical specification must reflect this analysis.

A good relationship with the manufacturing team is important to transmit the idea that QA/QC is not there to impede production; it is there, rather, to make sure that things go well and that the product is acceptable. Incredible as it appears, manufacturing contractors frequently perceive that measurements and checks slow production down: the result is that manufacturing may be of too poor a quality and needs to be redone, and painful extra costs are eventually generated by disassembly, corrective action and re-assembly.

On the other hand, the client often tends to ask for all possible measurements, just in case, without setting a proper data analysis process with clear acceptance criteria and often without proper models that can provide guidance on minimizing rejections and the length of production stoppages.

In the case of the LHC dipoles, the technical specifications [17] contained all checkpoints, the data were transferred to CERN on-line for immediate analysis and a team of analysts was always available to give an answer in half a day to a few days for the most difficult cases.

Also, resident inspectors, provided by a professional QA/QC company through a global contact for the whole LHC project, were placed by CERN at each production site, complementing the numerous visits of CERN technicians, specialist engineers and project engineers.

The most important measurements, apart the electrical ones, have been the coil size and the magnetic measurements. The coil azimuthal size allowed the determination of the thickness of the shims that are placed between coil and collar at the pole position (about 60°), see Fig. 4. The coil size varied typically ±50 μm, with some systematic fluctuation. This almost always allowed the use of nominal shim values, giving a big advantage for the uniformity of the harmonic content.

In Fig. 15 the values of the shim thicknesses used for the whole of production are reported [18]. The undulation of each single coil has been contained in 20 μm root mean square, a very good value for 15 m long coils with uniform shims.

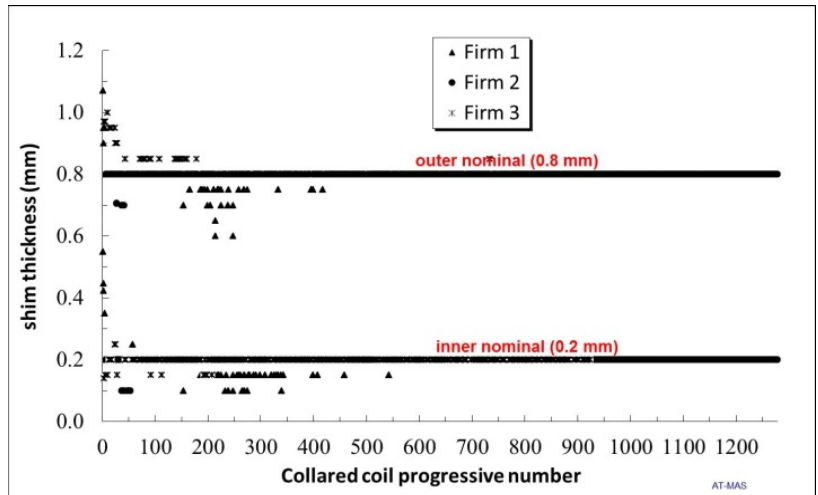

**Fig. 15:** Thickness of the shims placed between coils and collar pole, determining azimuthal extension and stress upon the coils.

As part of the QA plan, magnetic measurements were introduced both at the CC level and after completion of the magnetic circuit with the iron yoke. At the beginning they were thought mainly to steer production toward the beam dynamic target, i.e. to guarantee that the bending strength per unit current, $B_1 \times L_{magnetic}/I$ (also called the integrated transfer function), the sextupole $b_3$, the decapole $b_5$ and the septupole $b_7$ were inside the tolerance band. Two small fine-tunings of the cross-section, via changes of the copper wedge profile and thickness, have been necessary to maintain $b_3$ and $b_5$ during production in the desired limits, as shown in Fig. 16 for the sextupoles [18]. Magnetic measurements were revealed to be a powerful tool for finding, by means of a magnetic model of the coils, the exact position of a single cable or a coil block inside the CC when visual inspection and other methods were no longer possible. A detailed account of the errors and imperfections that have been intercepted with this method is reported in [19]. The accumulated number of important defects on the CC (requiring a coil de-collaring and corrective actions), intercepted through magnetic measurements, through dipole production showed a high rate of faults during two periods. The first period is reasonably related to the learning phase of the process when passing from prototype to production. After that initial phase, the production goes rather smoothly without any large problems, until a certain point when the fault rate has a steep increase. Later on, the sudden increase was correlated with the introduction into the workshop of new people, hired to staff the new production lines that had been installed to meet the required dipole production rate (see the following section on learning curves).

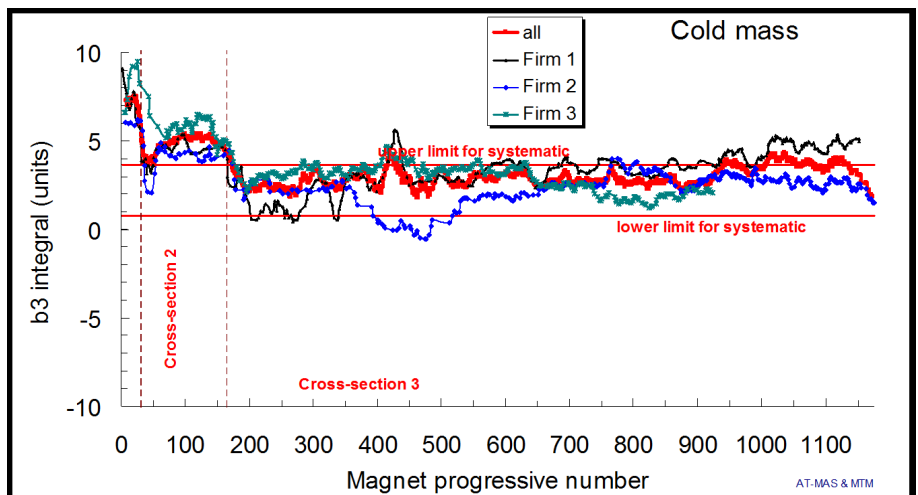

**Fig. 16**: Evolution of the normalized sextupole, b3, of the main LHC dipoles during construction. The vertical lines indicate the points when the two fine-tunings of the coil cross-sections were introduced.

## 5.2 Learning curves

Dipole production has been analysed in terms of so-called 'learning curves' [20]. A learning curve represents the production time for a unit when expressed as a percentage reduction of the production time for a unit when doubling the production number. When producing the first $N$ units, a time $\Delta t$ is required. A learning percentage ($\rho$) reduction of 90% means that to produce a further $N$ units it takes $0.9\Delta t$; then to produce a further $2N$ units a time $0.9 \cdot (1 + 0.9)\, \Delta t$ is necessary, and so on. This reflects an exponential decrease that can flatten very quickly for exponent near unity. In Fig. 17 we report the behaviour for the construction of the dipole for one of the three production lines. Table 2 indicates where LHC production is situated inside the panorama of industrial products.

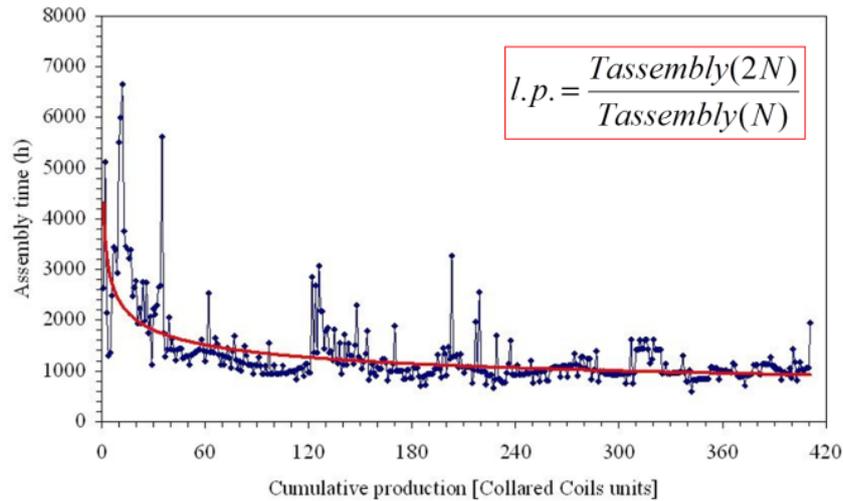

$$l.p. = \frac{Tassembly(2N)}{Tassembly(N)}$$

**Fig. 17:** Learning curves for LHC main dipole CC in one production line. The fitted line indicates the learning percentage (*lp* or $\rho$).

**Table 2:** Learning percentage ($\rho$) for LHC main dipole production compared with those of various other production items.

| Industry | Learning percentage ($\rho$) (%) |
|---|---|
| Complex machine tools for new models | 75–85 |
| Repetitive electrical operations | 75–85 |
| Shipbuilding | 80–85 |
| LHC magnets | 80–85 |
| RHIC | 85 |
| Aerospace | 85 |
| Purchased parts | 88–88 |
| Repetitive welding operations | 90 |
| Repetitive electronics manufacturing | 90–95 |
| Repetitive machining or punch-press operations | 90–95 |
| Raw materials | 93–96 |

It is interesting to examine the graph of the production forecast for single coils for one particular production line. Coil manufacturing may be the most delicate operation, heavily relying on the skill and training of personnel. The green (upper) curve in Fig. 18 expresses the projection of the company at the time of the plan (and costing forecast). The red and blue curves show the reality. It is interesting to note the correlation, done rather late in production, between the spikes indicating a sudden increase in the unit production time and the injection of new personnel into manufacturing, almost a textbook case. This shows the weight of the human factor in SC magnet production. Its success does not depend on design alone: proper tooling, a well-defined process and, above all, experienced or well-trained personnel play a major role in the quality and rate of production.

## 5.3 Industrialization and construction strategy

LHC dipole production can be taken as an example of an industrialization strategy based on a non-negligible investment with a very good result. Starting at the end of the 1980s, a relatively small, albeit slowly increasing, continuous program took place in close connection with industry. It may be legitimate to retrospectively wonder whether CERN went to industry too early. However, this helped to keep industry in the game and to gather support for the LHC from the CERN member states, by evoking the large industrial return that the LHC magnets entailed (in terms of money and technology transfer).

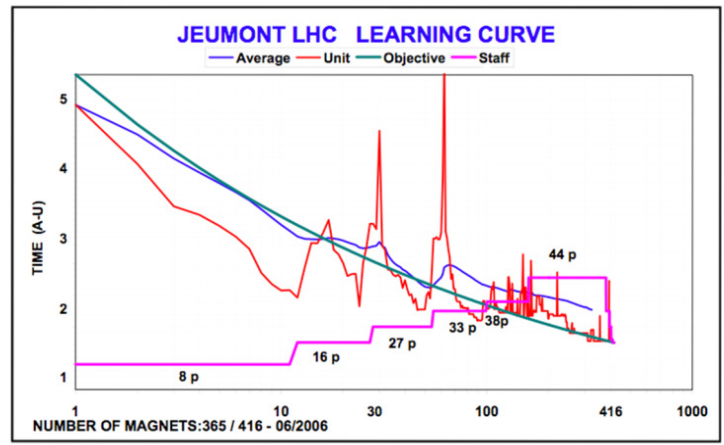

**Fig. 18:** Evolution of time needed for coil manufacture for LHC dipoles in one line (Alstom–Jeumont consortium), and its correlation with the introduction of new personnel.

Companies manufacturing LHC main dipoles were: Jeumont–Schneider and Alstom, later the Alstom–Jeumont consortium in France; Ansaldo Componenti, later Ansaldo Superconduttori Genova and then ASG in Italy; and Noell Pressaug, later Babcock Noell Nuclear and then BNG in Germany. An early involvement of Austrian Elin was discontinued in 1994 after a short model (in Nb$_3$Sn) and a long prototype; and an attempt to involve Oxford Instruments (UK) did not go beyond a short model. In Fig. 19 the first LHC dipoles, designed with a 50 mm bore, aluminium collars and 10 m long cold mass and developed in the CERN–INFN collaboration [21], is shown on the CERN test bench in 1994.

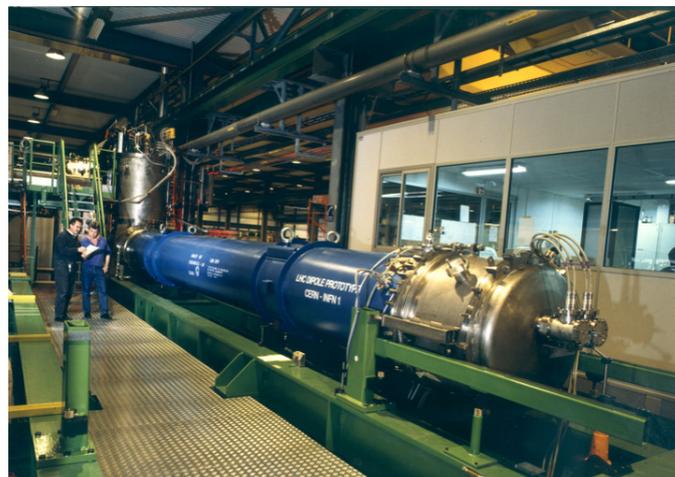

**Fig. 19:** The first LHC main dipole, 10 m long, on the test bench at CERN in June 1994, during the successful test campaign.

The other large magnets were fabricated without a long R&D phase between CERN (and collaborating institutes such as CEA) and industry. The company Accel in Germany was contracted to manufacture the series production of the main quadrupoles and part of the insertion quadrupoles. The remainder of the insertion quadrupoles have been manufactured by Tesla Engineering (UK). The special low-$\beta$ quadrupoles have been provided as an 'in-kind contribution' by Fermilab (US), who manufactured the magnets in its own premises, and KEK (Japan), who, after an R&D phase carried out in the laboratory, contracted construction to Toshiba.

Coming back to our story of the main dipoles, CERN kept the three companies in the business by continually assigning contracts: also, due to limited resources, CERN renounced manufacture of long magnets on its own site at the beginning. However, around 1995–1996 it became clear that at least one production line was required for cold mass completion in-house (CERN building 181, later called the CERN Magnet Assembly Facility (MAF)) to set up the many technical details.

As discussed above, these details may be less attractive or challenging from a technological point of view, but they were all needed to carefully set up and consolidate the following: insulation of the cold bore tube, the iron assembly method, pole and aperture connections, lyra and bus bars, shrinking shell welding technique, end-cover technology, corrector assembly, the curvature formation technique, alignment metrology, etc. These are the most critical from the list of technologies that had to be developed or set up in the CERN assembly facility.

A picture of CERN MAF around the year 2000 is given in Fig. 20. In 2001, with all work transferred to industry for dipole mass production, the CERN MAF was converted to an assembly hall for the cold mass assembly of the LHC insertion magnets.

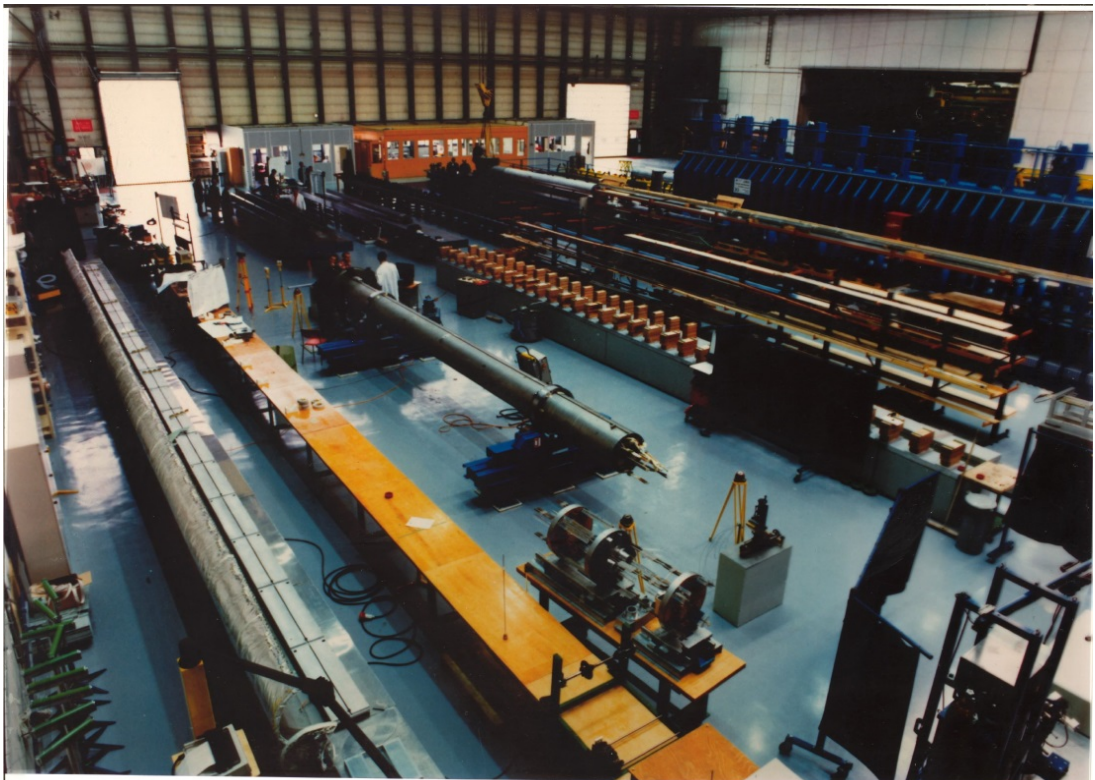

**Fig. 20:** CERN MAF (building 181, formerly the ISR experimental hall), with all equipment for dipole cold mass assembly around year 2000.

The strategy for procurement of the main components of the dipoles underwent a slow, but steady, change between the early years (1988–1992) and the final contracts for series construction, issued in 2001. In the initial years CERN was ordering turn-key finished magnets, completed in their cryostats, very much like HERA dipoles. However, for many reasons this idea had to be changed to arrive at a point where all main components were procured by CERN and supplied to the magnet manufacturers. In certain cases CERN procured even the raw materials, such as the low-carbon steel (iron) for the yokes, which was then given to various manufacturers for formation of the laminations via fine blanking; finally, the laminations were distributed to the various companies in charge of the construction of the magnets (dipoles and quadrupoles). CERN became a supplier to its own suppliers, but not without risk.

We can summarize the main advantages below.

- *Advance purchasing*. CERN could purchase critical components with a long lead time for procurement well in advance of placing the contract for magnet manufacture. A typical example is the low-carbon steel for yoke laminations, purchased at the end of 1997 just at the second and final approval stage of the project (the main magnet contracts were assigned in 1999 and 2001).

- *Economies due to large-scale purchasing*. By purchasing the quantities of materials for all of the dipoles (and in certain cases for quadrupoles too) better prices were obtained. The economy applies also to the component qualification process and follow-up checks (QA/QC). It is difficult to quantify the saving, because it also includes overhead savings that were also associated with a further step in purchasing, if components had been included in the main tender: probably, however, it is a considerable fraction (30–40%) of the total cost of components (~500 million CHF).

- *Technical compliance and uniformity*. It is always difficult to able to specify in a clear and objective way all of the characteristics that are actually needed in a component. We have examples of the fact that by assigning the components to a magnet manufacturer, although with an apparently detailed specifications, the magnets ended up with deviations from the range of acceptability without a real breach of the specifications (quench heater steel from a dipole manufacturer, beam tube fillet welding from a quadrupole manufacturer, etc.).

- *QA/QC on components*. It is almost impossible for different purchasers to have the same QA/QC and a follow-up with the same intensity and criteria of selection. No QA can substitute for skill and attentive inspectors. With CERN purchasing all components we ensured a homogeneity of characteristics that is very useful for series production, where uniformity of characteristics is in practice very important to avoid mistakes and non-conformities. When we had a deviation from specification we could evaluate it correctly, without suspicion that the non-conformity was the emerging part of a submerged iceberg, and we could also accept important non-conformities (for example, in the SC cables, in the collar profiles, etc.) because we could convince ourselves that such non-conformities would not adversely affect the final performance.

- *Security of supply*. By keeping in hand the procurement of components we could anticipate problems of quality, see above, but also of shortages. For example, we could avoid being trapped by a big strike in the steel foundry (the unique supplier of all low-carbon steel for the whole project), by removing all raw iron plates that were in stock there with an exceptional transportation arrangement (about 400 trucks were organized in a real blitz) to a temporary store under our control. In another case we helped to start up a new company for production of critical components (quench heaters). We could feel that production from one of the two production centres was of increasingly bad quality and we anticipated its stoppage due to a lack of profit. The new company then became the required second production centre for quench heaters. All of these actions could have hardly be seen by the magnet manufacturers and it would have been very surprising to see them acting as we did

(a supplier is always protected by the clause of *force majeure* in the case of a national strike, or of a sudden stop by a subcontractor).

- *In-kind contributions and contract distribution.* CERN is an international organization financed by 20 member states (MS); and for the LHC project various non-member states (NMS) made an important contribution. By keeping component procurement in-hand CERN could spread in an optimal way the distribution of the contracts among MS, thus favouring a balanced return, without however compromising on the technical aspects and without spending more than the minimum (with a very few exceptions, such as upon a contract for iron laminations). If global contracts, including tooling and components, had been assigned to magnet manufacturers, making these a kind of general contractor, this balance would have been much more difficult to reach and certainly less optimized. On this line alone the centralization of procurement could make the best use of the in-kind contribution by the NMS. For example, supplying the Nb–Ti and Nb sheets, from the US, to superconductor manufacturers is probably an absolute 'prima' in the panorama of superconductor production, impossible without centralization and an assumption of responsibility by CERN. Other very successful examples are the supply of all polyimide insulation foils and tapes for all coils, as well as the supply of austenitic steel for the collars for all dipoles: both produced by Japanese companies and procured by CERN for the various component or magnet manufacturers in Europe.

The list of the inevitable drawbacks associated with such a strategy is given below.

- *Liability for delays.* This is the first price to pay: make sure that delivery is on time or at least 'just in time'. Many times we really went to the borders of a just-in-time deadline; however, for the dipoles we managed to avoid any stoppages (which would have cost about 1.5 million CHF per week). We had one case of large delays for the main quads (corrector magnet supply): despite all actions an extra cost of several million CHF had to be paid, a big number in absolute terms, but small compared with the total savings. One has to say that as the final client we could not apply to ourselves, or toward our magnet manufacturers, the clause of *force majeure* that contractors have toward the client.

- *Additional workload for procurement.* The CERN Magnet group organized a Superconductor section and a Components Centre section, which together with the purchasing office took care of the procurement for the SC cables and the other components (a few other components were procured by the Magnet Laboratory and Insulation section and by the Cryostat and Integration group, later merged with the Magnet group). Interfaces between CERN and magnet manufacturers became more complex and we had to manage together the components and the magnet manufacturing schedules. All of these could be evaluated in an increase of personnel to follow-up the components: about six full-time equivalents (FTE) × 5 years for SC and about four FTE × 5 years for the Component Centre, i.e. about 50 FTE-years, excluding the Procurement office and Finance office personnel.

- *Technical responsibility and interfaces.* Each important component was specified by CERN. However, with procurement CERN became fully responsible for their technical quality, too, with respect to assembly problems or performance risk. Clearly distinguishing whether a failure is due to a component imperfection rather than a non-conforming use of it by the magnet manufacturer is not trivial, at all, in many cases. Examples of such interfaces were especially numerous in coil construction: short circuits may have been caused by burrs and chips on the copper longitudinal wedges or defects of the insulating tapes (both supplied by CERN) or by metallic inclusions during winding in a non-proper environment.

- *QA documentation responsibility.* By supplying components CERN also become entangled in the each company's QA plan, which had to be fed by the component data. A delay in QA documentation would have become a cause of trouble to the company. In a way, this entangled situation forced CERN to become very efficient with the QA plan to avoid incurring penalties (which happened in reality, fortunately however at a very minor level). On the other hand CERN could be strict in imposing the respect of the QA plan by the magnet manufacturers, having shown its good will in this respect.

- *Logistics*. The extra workload also meant managing advanced logistics. Not only did CERN have to work with a coordinated plan, being careful in not favouring/disfavouring any of the magnet manufacturers, but it had to organize the transport, storage and the QA documents' availability for a great number and variety of components. In total 150,000 tonnes of materials were transported all over Europe (and from North America and Asia). For more than 4 years we had to follow at least five large trucks per day in Europe, having reserves for any inconvenience, including stolen materials (which also happened, of course, involving precious components such as the SC cable).

Altogether the success of the strategy was overwhelming. Of course this meant that any lack of performance of a magnet at acceptance was not necessarily due to bad manufacturing. CERN was not only responsible for the design and assembly procedure, but also for components, whose behaviour is critical for the magnets' performance. We want to underline here that it is clear, from what has previously been said, that industry has been very cooperative, an attitude to be credited to the quality of the industry involved and that could be gained and reinforced thanks to the long R&D and industrialization phases, during which a strong, mutual trust between CERN and industry was built.

The path from reaching 9 T in a single short dipole, which happened in 1987, and from the success of the first 9 T long prototype, which happened in 1994, to the point where design and technology was mature enough to contract construction of a large number of 8–8.5 T operational field full-size dipoles, which happened in 2002, was longer than expected. In comparison, the period of construction was relatively short, despite the large number of units: the last dipole and last quadrupole were delivered to CERN on 7 November 2006. The time for installation and commissioning, also because of the incident of September 2008 [22] that plagued machine start-up, was also longer than expected, resulting in full operation of the LHC from 30 March 2010. The LHC magnet timeline is reported in Fig. 21.

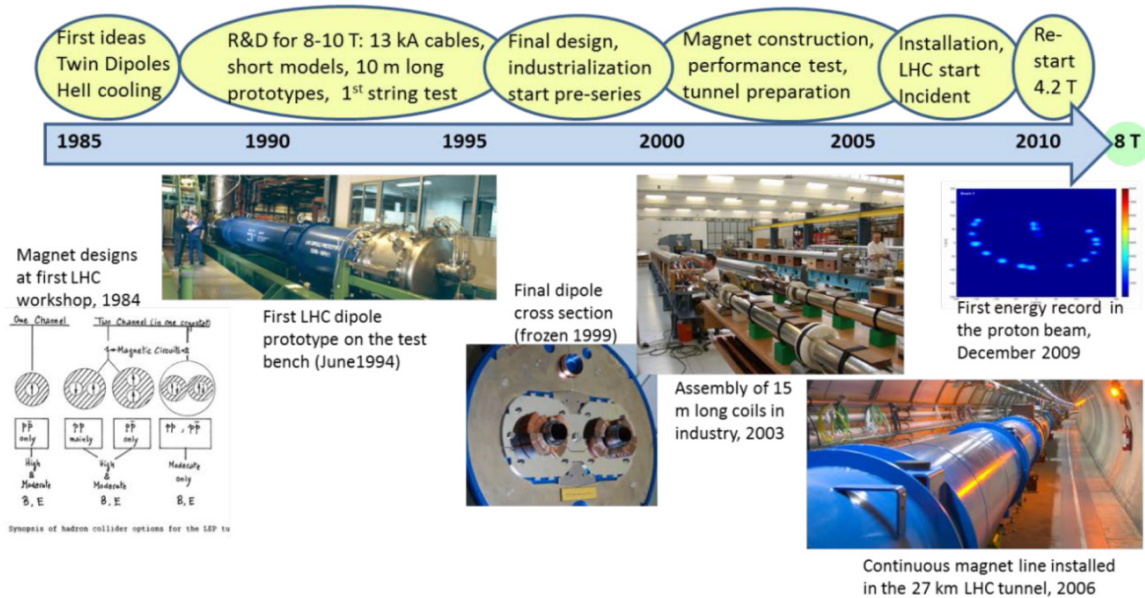

**Fig 21:** LHC magnet time line

In this paper we do not mention the specific task of cryostat manufacture. In comparison with magnet construction the cryostat is mainly a thermo-mechanical construction, with far fewer problems of tolerance and accuracy. However, the cryostat is the point of the integration of the magnet in the cryogenic system and the interface with the electrical circuit as well as with the other services. This complexity needs to be discussed in the early stage of the magnet design, taking the necessary feedback, and requires a close collaboration between the magnets' designers and cryostat engineers. In the end, it is not the magnet that has to work correctly, but the magnet system in its complexity that is more by far than that of a single magnet.

# 6 Novel construction technology: Nb$_3$Sn and high-temperature superconductors

So far all existing accelerators have employed Nb–Ti-based magnets, whose construction issues have been discussed in the previous chapter. However, it is more than 20 years since studies were performed to increase the performance and to open the way for fields in excess of 10 T. To this aim a Nb$_3$Sn superconductor is necessary, of course, with all of the technology consequences entailed by this extremely brittle compound that necessitates a thermal treatment for its formation.

A non-exhaustive list of manufacturing issues when Nb$_3$Sn technology is applied to accelerator magnets is given below.

i. Use of *wind and react* technology (first winding the coil with a ductile conductor, then the entire coil is treated at 650–700°C to react the conductor, forming brittle Nb$_3$Sn) is mandatory because the coil radius of curvature is 10–50 mm which, with a cable of 1–1.5 mm thickness means more than 1–2% strain, much beyond the 0.3% safety limit and even the 0.6–0.7% irreversible degradation threshold.

ii. The wind and react route is very demanding: all coils components (mainly insulation, but not this alone) must withstand the 650°C reaction temperature, maintaining their properties and avoiding decomposition into carbon. A lot of information is available about various types of possible glass fibres that can be used: S-glass or, better, its non-military grade, less expensive, S2-glass variant, or the similar European standard R-glass. The more common E-glass could be used, but its annealing point is dangerously close to 650°C and caution must be taken in the presence of high stresses. Glass tape or braid being a weave, from the electrical point of view behaves basically like an inter-turn spacer; solid insulation is actually ensured by the resin impregnation (usually under vacuum), which is a critical operation for electric, mechanical and thermal properties. Glass-epoxy can be reinforced with mica-glass composite tape. Choice of the sizing agent for the glass is also critical to avoid excessive carbon formation. A mild grey coloration of the reacted coil is generally acceptable (carbon increases its resistance by a factor ~1000 when cooled from 300 to 2 K).

iii. Coils and their tooling should allow for expansion–contraction over almost 1000°C (from 300 to 970 K and then down to 4 K) while keeping an accuracy of 30–50 μm. The SC cable after Nb$_3$Sn formation at 650°C shows a completely different behaviour in terms of thermal expansion and this must be taken into account to avoid over-stressing. The actual behaviour is strongly dependent on the copper content and on the particular process used to obtain Nb$_3$Sn: bronze route, classical internal-tin-diffusion, rod-restacking process, powder-in-tube, etc.

iv. The degradation of the critical parameters of Nb$_3$Sn due to strong transverse stress is almost unavoidable and a large effort in conductor development, coil technology and magnet structural design is necessary just to mitigate this effect. Of course this is not easy because Nb$_3$Sn is used for high fields, $B > 10$ T, when the force and stress build-up become considerable.

v. The field quality (and sometimes quench performance, too) is affected by large filaments, very large $J_c$ at low field, low conductivity in the stabilizer, large intra-cable coupling currents, etc. These effects requires mitigation measures such as, for example, a high-resistance core inside the Rutherford cable, in the form of stainless steel tape, which affects the mechanical and stability properties of the conductor and winding.

Use of $Nb_3Sn$ for accelerator magnets is definitely not straightforward. The Conductor Development Program of DOE (CDP-DOE) launched in the US in 1998 [23] was aimed at the following specific goals for building an accelerator much beyond the existing $Nb_3Sn$ then available (1995), for fusion and high-field solenoid programs: i) $J_c$ of at least 3000 A/mm$^2$ at 12 T, 4.2 K was the primary goal, which was attained quite soon; ii) then an effective filament size of 30–50 μm (for field quality) and high Residual Resistivity Ratio (RRR) in the stabilizer became the main goals and this is still work in progress.

Considering construction technology, a number of companies are able to produce $Nb_3Sn$ solenoids (for high-field laboratory magnets and NMR spectroscopy); a few specialized companies are accomplishing the task of manufacturing the huge $Nb_3Sn$ coils for the ITER project; however, no company today possesses the technology for manufacturing $Nb_3Sn$ accelerator magnets. It is one of the challenges for the next project, High Luminosity LHC, to be able to set a mature and solid $Nb_3Sn$ technology for accelerator magnets while starting the necessary technology transfer to industry.

In Fig. 22 a mosaic of the tooling and coil construction at various stages is shown for the $Nb_3Sn$ 11 T dipole project (Fermilab–CERN collaboration) [24]. The 11 T dipole is designed to use a 'classical' collar structure for force containment. In the context of the large-aperture quadrupole R&D for LHC Accelerator R&D Program (LARP), a US program oriented to the High Luminosity LHC upgrade [25], and for their own internal high-field dipole program, Lawrence Berkeley National Laboratory (LBNL) designers have devised a new concept called 'key and bladder'. Here, collar function is reduced to a minimum: they are very thin and used just for coil assembly.

The restraining forces are generated by an external solid shrinking cylinder, via inflation of high-pressure bladders. The insertion of a suitable key maintains the pre-stress when the bladders are de-pressurized and finally removed. The restraining forces are smaller than those usually applied by a collar, but they increase during cool-down because the aluminium cylinder contracts more than the $Nb_3Sn$ coils. More details can be found in [13, 26].

Clearly this technology can also be applied to Nb–Ti coils. However, it is particularly interesting for $Nb_3Sn$ magnets, because coils wound with this conductor have a higher (two to three times) elastic modulus than coils made out of Nb–Ti; in addition, $Nb_3Sn$ coils are also less dimensionally accurate. A collar structure controls the coil size, and stress comes as a consequence (assuming collars to be of infinitely rigid structure), while a key and bladder structure controls the force, and coil size comes as a consequence.

Owing to the higher coil rigidity and reduced dimensional accuracy, collar structures for $Nb_3Sn$ windings may result in excessively high pre-stress. Such a dangerous situation is much less probable in the case of a key and bladder structure. However, this is a new technology that has still to be demonstrated in long, operational quality, magnets.

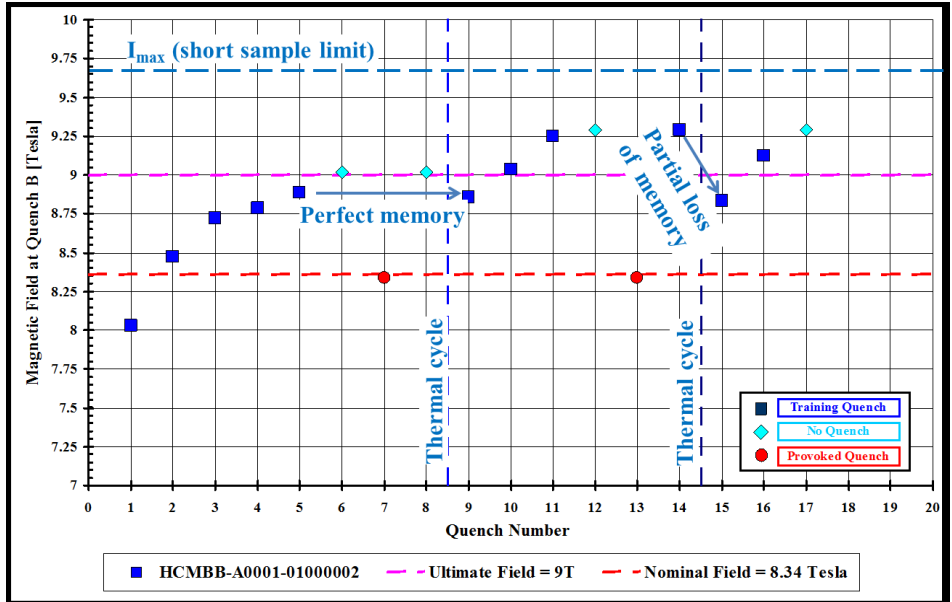

**Fig. 22:** Quenches of LHC dipole 1002 at virgin training curve and after the first and second thermal cycles (indicated by vertical lines).

## 7 Testing

A SC magnet that has not yet been tested is worth virtually nothing. The scope of magnet testing radically changes between the R&D phase and the construction phase. During R&D the scope of the test is to assess the limit of the design and the merits of the various choices, and assess the best technology in terms of performance and reliability.

In the construction phase, and particularly with a massive production series, the test has two main goals: i) to serve as ultimate QA tools, to establish whether the whole magnet system is performing as specified: quench performance is important in this respect but it is 'only' one of the many important aspects of the system's performance; ii) to serve as an acceptance test, with all of its financial and legal implications that a laboratory has toward the manufacturer.

Here we will briefly discuss tests only in relation to validation of design and construction: a review of the LHC magnet test can be found in [27].

### 7.1 Test for series production

Magnet testing in operating conditions is the ultimate QA tools because not only is it necessary to ascertain whether the work was done well: it is also necessary to intercept various mistakes, such as defects in components, fabrication errors and design faults or weaknesses that cannot be evidenced in any way other than the final test in full operating conditions. Examples of such weak points are: a too high resistance in the SC joints; premature quenches due to mis-positioning of the SC cable; wear of insulation because of cool-down induced movements, and so on. The importance of testing as the ultimate QA measure is evidenced by the fact that it may, and usually does, trigger corrective actions in the manufacturing chain in order to remove, or at least mitigate, the mistake and other sources of non-conformity. Therefore, testing in operating conditions should really be part of the production cycle. The promptness of the feedback must be proportional to the production rate: taking the LHC as an example, when each company was at the maximum production rate of three dipoles per week, the 3-month contractual delays between magnet delivery and testing meant that 40 dipoles were produced between the eventual discovery of a systematic defect and the implementation of the corrective actions in the production chain. However, 40 dipoles are almost 10% of the whole production: readiness in testing is one of the best ways to avoid huge extra costs and delays in production.

The result of the test as an acceptance test is to pay the manufacturing company upon positive results, and to release it from any further responsibility, except of course for hidden defects, for which should be agreed a clause of responsibility lasting for a few years, enforced by a solid bank guarantee. A positive acceptance test means that the magnet is qualified for operation. The temptation to understand the design limits, by pushing magnets beyond what is needed for operation, should be resisted because it is mixing acceptance testing and the R&D phase. Once a magnet is accepted by the group/team clearly in charge of it, a further analysis of the magnet's properties, to verify the quality of the magnet and decide the proper installation slot, may be demanded by a suitable body. For the LHC this was the Magnet Evaluation Board, which was off-line from the magnet acceptance chain.

### 7.2 Testing during the magnet research and development phase

In the R&D phase the magnet is pushed beyond the operating conditions, to learn about design choices and actual performance limits and, therefore, operational margins.

Before describing the test it is useful to explain that accelerator magnets, when pushed near the critical current limit and/or the force-retaining structural limit, have a typical behaviour called 'training'. Quench happens at a level of current that is lower, and in certain cases much lower, than the maximum theoretical current, which is the critical current along the load line, also called the short sample limit because it is inferred from measurements of a short cable sample cut from the extremities of the unit length. Assessing training and limits is one major goal of testing an accelerator magnet.

The test starts with a cool-down that must be carried out in a well-thought-out manner, with an acceptable temperature gradient to avoid excessive thermal stresses. The electrical insulation is then checked: this check has to be done at each phase, starting with reception from the manufacturer, at room temperature. Electrical checks are critical to assessing magnet integrity and must be really well thought out, at a voltage that should slightly decrease between successive tests. The protection circuit must then be tested at a low current. Finally, the power test can proceed, by raising the current with a ramp rate equal to or similar to that in operation.

Usually the magnet undergoes a sudden transition (quench), losing SC status at a value of current well below the design one. The current is quickly removed (100 ms), however the magnet heats up in 1 or 2 seconds, reaching ambient temperature in a hot spot. The magnet is cooled again and re-energized until another quench is recorded at a value of current usually higher than the previous one: it looks as though the magnet is being 'trained' toward higher values. The sequences of quenches are called training curves, and a typical sequence is reported in Fig. 22. After the training the magnet reaches flat values. The eventual difference between the target value and the flat value is called degradation (more than 5–10% would be considered a serious fault of design or fabrication). Usually the flat value is above the operating conditions, as in Fig. 22, since the magnet is actually designed to have some margin, to account for shortfalls and degradation during series construction. Referring to Fig. 22, it can be seen that at the LHC we defined an operational target (nominal field value, 86% of the short sample limit) and also an ultimate field value (93% of the short sample limit), corresponding to the limit of mechanical design, bus bar system, connections, etc.

One further important performance parameter is the so-called 'memory'. After the training curve, the magnet is completely warmed up (the warm-up induced by quench alone is never complete). The magnet is cooled down again and is ready to repeat a number of powering-to-quench cycles. This is the second training curve. One can then repeat the thermal cycle and have a third training curve. It can be seen from Fig. 22 that this dipole magnet, actually one of the best of the pre-series, is well qualified for its operational field and is marginally good for its ultimate field: 9 T is reached and also passed at the second and third training curves; however, some loss of memory happens around its ultimate value. The implications of commissioning accelerator operation go beyond the scope of this paper. Here it suffices to say that some of the LHC dipoles exhibit a small loss of memory around the nominal value. This will probably imply that the nominal value in the LHC (8.33 T, corresponding to a 7 TeV/beam) will be achieved later in the future because, even if only 30% of the LHC shows some loss of memory, this means that some 400 quenches will be required in the tunnel during re-commissioning of the LHC in 2014–2015. It may well happen that, if the number of re-training quenches is greater than our expectation, the LHC main dipoles may be limited to working at around 80–83% of their short sample value, as in all previous accelerators. This would mean limiting the beam energy to 6.5–6.8 TeV/beam.

Another important goal of the test is to measure the field and the harmonic content of the magnetic field in the region used by the beam. This property is usually called field quality. A good part is usually measured at room temperature, i.e. at a very low current and without SC effects, as mentioned in the Section 5.1. Usually these measurements give a very good prediction of the field quality due to the geometry of the conductors. However, measurements in operational conditions, at cold, need to be done: i) to assess the exact form of the warm–cold correlation for geometric field quality; and ii) to assess the behaviour of the SC effects. For the LHC a large number of studies allowed changing the strategy for magnetic measurements in an advantageous way: also, due to pressure posed by a delay in testing, the cold field measurements have been reduced from an initial 100% down to about 15% (see [18, 28]). It is worth saying that the excellent field quality of the magnets, and complete knowledge of its operation, are some of the most important ingredients of the great success of the LHC in its first years of operation.

# 8    Conclusions

Manufacturing a SC magnet of accelerator quality is a unique challenge. As good as the design may be, a magnet that is manufactured badly with low-quality or bad components and tooling will perform poorly, impeding a correct evaluation of the design merits. In this paper we have tried to summarize the main principles of accelerator magnet construction, and to recollect the main lesson gathered from the LHC adventure, underlining the close tie among design, manufacturing process, tooling and testing. Superconductivity calls for a total quality process: different from room-temperature technology, even an apparently minor detail can have detrimental effects on the magnet's performance. In this respect the engineering design and the construction and testing represent a true intellectual challenge, especially in large projects where resources are limited and great pressure is put on schedules. The main lessons we can draw from LHC magnet construction can be summarized in two points:

i. one must not compromise on quality and on controls as any hidden defects will show up, at the worst possible time;
ii. even more importantly, especially when resources are limited as is the case in all practical projects, 'the best is the enemy of the good'; when a solution is sufficient to the scope, one has to resist any improvements since, in a complex system such as a SC magnet, a change in one part will likely affect, sometimes in a subtle way, the whole system.